\documentstyle[preprint,aps,epsf,floats]{revtex}
\tighten
\def\usss{\Upsilon(3S)}
\def\uss{\Upsilon(2S)}
\def\us{\Upsilon(1S)}
\def\pss{\psi(2S)}
\def\ps{J/\psi}

\begin{document}

\preprint{DOE/ER/40427-31-N96}
\title{Hadronic and Electromagnetic Interactions of Quarkonia}
\author{Jiunn-Wei Chen 
\footnote{{\tt jwchen@phys.washington.edu}}
and Martin  J. Savage 
\footnote{{\tt savage@phys.washington.edu}}
 }
\address{
Department of Physics, University of Washington, 
\\
Seattle, WA 
98195
}
\maketitle

\begin{abstract}

We examine the hadronic interactions of quarkonia, focusing on 
the decays $\pss\rightarrow\ps\pi\pi$ and
$\uss\rightarrow\us\pi\pi$.
The leading gluonic operators in the multipole expansion are matched
onto the chiral lagrangian with the coefficients fit to  available data,
both at tree-level and loop-level in the chiral expansion.  
A comparison is made with naive expectations loosely based on
the large-$N_c$ limit of QCD in an effort to  
determine the reliability of this limit for other observables, 
such as the binding of $\ps$ to nuclei.
Crossing symmetry is used to estimate the cross-section for
inelastic $\pi\ps\rightarrow\pi\pss$ scattering, potentially
relevant for heavy ion collisions.
The radiative decays 
$\pss\rightarrow \ps \pi^+\pi^-\gamma$ and 
$\uss\rightarrow \us \pi^+\pi^-\gamma$
are determined at tree-level in the chiral lagrangian.
Measurement of such decays will provide a test
of the multipole and chiral expansions.
We briefly discuss decays from the $\usss$ and also the contribution
from $\pi$'s to the electromagnetic polarizability of quarkonia.

\end{abstract}

\bigskip
\vskip 4.0cm
\leftline{October 1997}

\vfill\eject

\section{Introduction}

The strong interactions of quarkonia, such as the $J/\psi$ and $\Upsilon$ 
are much simpler than  those of other hadrons as they are
compact on the hadronic scale.
Gluons with a wavelength $\lambda$ much larger than the ``radius'', 
$r_{Q\overline{Q}}$, of quarkonia
have  interactions suppressed in a multipole expansion
by powers of $r_{Q\overline{Q}}/\lambda$.
The leading operators involving 
two gluons correspond
to interactions with external chromo-electric 
and chromo-magnetic fields
\cite{Gota,VZa,Volb,Peska,BhanPesk,NovShif,KYa,YYa},
with 
higher order interactions naively suppressed by additional 
factors of the quarkonium radius.  
In order to determine the interactions of quarkonia with other 
hadrons nonperturbative matrix elements of the gluonic operators 
between the hadronic states must be evaluated. 
In general, such matrix elements cannot be determined simply, 
other than by lattice computations.
However, there are a few special and well known cases where the 
matrix elements of certain gluonic and light quark operators are 
known by the symmetries of QCD.

There has been renewed interest in the interactions of the lowest 
lying hadrons with quarkonia.
An exciting prospect is measuring the binding of $J/\psi$ to nuclei
\cite{BSTa,LMSa,KVa}, 
or even measuring the scattering cross section from a single nucleon
\cite{BMa}, but theoretical 
computations require knowledge of the couplings between 
gluons and the $J/\psi$.
A large-$N_c$ estimate of these couplings
\cite{Peska,BhanPesk}\ 
has been used to predict a binding energy of $\sim 10 \ {\rm MeV}$ to 
infinite nuclear matter.
Further, the observed anomaly in elastic 
polarized $pp$ scattering near 
$5\ {\rm GeV}$\cite{Cosa} has been attributed to threshold or 
resonant effects in $c\overline{c}$-N interactions\cite{BTb}.
The low-momentum interactions of quarkonia are also important in a
completely different arena.
It is possible that the ratio of observed to predicted $J/\psi$ 
production in ultra-relativistic  heavy ion collisions will provide 
a ``handle'' on the formation of a quark-gluon plasma in the collision
\cite{MSa}.
Whether or not this turns out to be valid, it is important to 
understand the interaction of quarkonia with the debris 
resulting from such a collision.  The existence of co-movers 
with the quarkonia
\cite{GVa} requires that the
low-momenta interactions as well as hard interactions be understood.

The natural framework for dealing with low-momentum interactions between 
the light hadrons and quarkonia is chiral perturbation theory.
This allows one to relate matrix elements for  different 
processes using the underlying symmetries of QCD.
Chiral symmetry has been used to constrain 
pionic matrix elements in previous work 
\cite{VZa,NovShif,BCa}
and further a chiral lagrangian in the meson sector was 
written down at leading order
\cite{GRa,MUa}.
This was used to determine the leading contribution of the 
light quark masses to quarkonium level 
spacings\cite{GRa}, and  to examine the decays 
$\pss\rightarrow \ps \pi^+\pi^-$,
$\uss\rightarrow \us\pi^+\pi^-$\cite{MUa}.

We explore several aspects of the interactions between 
quarkonia and the low-lying hadrons.
Firstly we wish to determine if statements can be made concerning the 
inelastic scattering of $J/\psi$'s by co-moving $\pi$'s as are 
to be expected in ultra-relativistic heavy ion collisions.
Secondly, we wish to test the predictions of the 
large-$N_c$ limit of QCD 
for the coefficients of the chromo-electric and magnetic operators.   
These directly impact the predictions for the binding of 
quarkonium to nuclei.
Thirdly, we wish to test the usefulness and validity of the 
chirally symmetric lagrange density 
that matches onto the multipole expansion.
To this end we explore the radiative decays   
$\psi (2S)\rightarrow J/\psi \pi^+\pi^-\gamma$ and 
$\Upsilon (2S)\rightarrow \Upsilon (1S)\pi^+\pi^-\gamma$.

\section{The Multipole Expansion}

The lagrange density describing the interaction between quarkonia 
and gluons at leading order in the gluon momentum and 
the heavy quarkonium limit is
\begin{eqnarray}
{\cal L} & = & 
\sum_v\ {1\over \Lambda_{Q\overline{Q}}^3}\ 
\left( P^\dagger_v (f) P_v (i)  - V^\dagger_{\mu v}(f)  V^\mu_v (i) \right)
\left( \tilde C_E^{(f,i)} {\cal O}_E \ +\  \tilde C_B^{(f,i)} {\cal O}_B\right)
\ \ \ \  .
\end{eqnarray}
The subscripts denote chromo-electric and -magnetic interactions respectively,
\begin{eqnarray}
{\cal O}_E & = &  -v_\mu v_\nu G^{\mu\alpha A} G^{\nu A}_\alpha\nonumber\\
{\cal O}_B & = &  {1\over 2} G^{\alpha\beta A} G_{\alpha\beta}^A\ 
-\ v_\mu v_\nu G^{\mu\alpha A} G^{\nu A}_\alpha
\ \ \ \ ,
\end{eqnarray}
and $v_\alpha$ is the four-velocity of the quarkonium.
The quarkonia field operators are normalized non-relativistically
(factors of $\sqrt{2 M}$ removed),
where $P$ denotes a pseudo-scalar quarkonium and $V$ a vector quarkonium
in the same multiplet of heavy-quark spin-symmetry.
The $i$ and $f$ that appear as superscripts on the constants 
$\tilde C_{E,B}$ 
and as labels on the field operators denote the initial and final 
states of quarkonia, e.g. $1S$, $2S$, etc.
By this construction we are implicitly assuming that the momentum 
transfer from the gluons is 
small, and the four-velocity of the quarkonium is 
conserved during interactions.
The mass scale  $\Lambda_{Q\overline{Q}}$ will be of order the inverse 
radius of the state,
and as such we expect the coefficients $\tilde C_{E,B}^{(f,i)}$ to be 
numbers of order unity.
In order to progress further in computing observables, these 
coefficients $\tilde C_{E,B}^{(f,i)}$
need to be determined.   As the wavefunctions for color-octet 
intermediate 
quarkonium states are unknown,
one must resort to models or particular limits of QCD for 
theoretical estimates.

For decays, such as 
$\psi (2S)\rightarrow J/\psi \pi^+\pi^-$,
$\Upsilon (2S)\rightarrow \Upsilon (1S)\pi^+\pi^-$
it was argued that the multipole expansion is not applicable as the 
energy release is the same order as the denominator in 
the multipole expansion
\cite{LSa}.
While it is true that there is no tuneable small parameter that 
can make the multipole expansion arbitrarily
accurate for such decays, it is possible that numerical
factors allow the expansion to be convergent.
This might arise from the fact that the intermediate states
that appear in the matching are in a color octet and will not 
be degenerate with the analogous color singlet states.  
The scenario appears to work well for the strong decays
$\pss\rightarrow\ps\pi\pi$ and $\uss\rightarrow\us\pi\pi$
as discussed previously and in the following sections.
However, in order to test this further we will discuss the 
radiative decays
$\pss\rightarrow\ps\pi\pi\gamma$ 
and $\uss\rightarrow\us\pi\pi\gamma$
which are calculable at leading order in 
the multipole and chiral expansions.

The gluonic operators ${\cal O}_{E,B}$ can be written in terms of the 
gluon energy-momentum tensor,
$T_g^{\mu\nu}$
\begin{eqnarray}
{\cal O}_E & = &  v_\mu v_\nu T_g^{\mu\nu}
\ -\ {1\over 4}  G^{\alpha\beta A} G_{\alpha\beta}^A
\nonumber\\
{\cal O}_B & = &  v_\mu v_\nu T_g^{\mu\nu}
\ +\ {1\over 4}  G^{\alpha\beta A} G_{\alpha\beta}^A
\ \ \ \ ,
\end{eqnarray}
where 
\begin{eqnarray}
T_g^{\mu\nu} & = & {1\over 4} g^{\mu\nu}  G^{\alpha\beta A} G_{\alpha\beta}^A
\ -\ G^{\mu\alpha A} G^{\nu A}_\alpha
\ \ \ \ .
\end{eqnarray}

Higher dimension operators such as those discussed in \cite{LSa} 
are suppressed by powers of the quarkonium radius, 
or possibly more appropriately the 
spacing between the color singlet and octet states.
We will not address the role of the higher dimension operators in the 
multipole expansion.
For  convenience we will define coefficients
$ C_{E,B}^{(f,i)}$ through
\begin{eqnarray}
C_{E,B}^{(f,i)} & = & {\tilde C_{E,B}^{(f,i)} \over \Lambda_{Q\overline{Q}}^3 }
\ \ \ ,
\end{eqnarray}
where the units of the $C_{E,B}^{(f,i)}$ are ${\rm GeV}^{-3}$
throughout this work.

The amplitude for a process, say $\pi\ps\rightarrow\pi\pss$ at leading order
in the multipole expansion
is then given by 
\begin{eqnarray}
{\cal A} & = & \langle \pi\pss| {\cal L} |\pi\ps\rangle
\nonumber\\
& = & 
(C_E+C_B)\  v_\mu v_\nu \  \langle \pi\pss| T_g^{\mu\nu} |\pi\ps\rangle
\ +\ 
(C_B-C_E)\  \langle \pi\pss| {1\over 4}  G^{\alpha\beta A} G_{\alpha\beta}^A 
|\pi\ps\rangle 
\ \ \ \ .
\end{eqnarray}

\section{Matching onto the Chiral Lagrangian}

To determine physical scattering amplitudes,  matrix elements of 
$T_g^{\mu\nu}$ and $G^{\alpha\beta A} G_{\alpha\beta}^A$ are 
required between the appropriate light hadron states.
For low momentum processes chiral perturbation theory 
is the natural framework to work.
The strong dynamics of the octet of pseudo-Goldstone bosons 
is described at leading order by a lagrange density 
\begin{eqnarray}
{\cal L}^\pi & = & 
{f^2\over 8} Tr\left[ \partial_\mu\Sigma \partial^\mu\Sigma^\dagger \right]
\ +\ 
\lambda Tr\left[ m_q\Sigma \ +\ h.c \right]
\ +\ ...
\ \ \ \  ,
\end{eqnarray}
where 
\begin{eqnarray}
\Sigma & = & \exp \left({2i\over f} M\right)
\ \ \ ,
\end{eqnarray}
is the exponential of the meson field (in $SU(2)_L\otimes SU(2)_R$) 
\begin{eqnarray}
M & = & \left(
\matrix{
\pi^0/\sqrt{2} & \pi^+ \cr
\pi^- & - \pi^0/\sqrt{2} }
\right)
\ \ \  ,
\end{eqnarray}
and $m_q$ is the light quark mass matrix.
The dots denote higher dimension operators suppressed by inverse powers of the 
chiral symmetry breaking scale, $\Lambda_\chi\sim 1 {\rm GeV}$.

\subsection{ Matrix Elements of $ G^{\alpha\beta A} G_{\alpha\beta}^A$}

It is well known that the divergence of the scale current 
allows one to determine matrix elements of the 
$ G^{\alpha\beta A} G_{\alpha\beta}^A$ operator
\cite{VZa,Volb,NovShif,LMSa,CCGGM} at low momentum transfer.
The divergence of the scale current (the trace of the 
energy-momentum tensor) for QCD is 
\begin{eqnarray}
\partial_\mu s^\mu & = & 
{1\over 2} {\beta(g)\over g} G^{\mu\nu A} G_{\mu\nu}^A 
\ +\  \sum_i (1-\gamma_i) m_i\overline{q}_i q_i
\ \ \ \ ,
\end{eqnarray}
where the $\gamma_i$ are the anomalous dimension of the 
$\overline{q}_i q_i$ operator and the $\beta$ is the QCD
$\beta$-function, 
\begin{eqnarray}
\beta (g) & = & -{9 g^3\over 16\pi^2}
\ \ \ ,
\end{eqnarray}
for $3$ flavours of light quarks.
For our work we will neglect the $\gamma_i$, as they are 
small compared with unity and 
work with 
\begin{eqnarray}
\partial_\mu s^\mu & \rightarrow & 
{1\over 2} {\beta(g)\over g} G^{\mu\nu A} G_{\mu\nu}^A 
\ +\  \sum_i  m_i\overline{q}_i q_i
\ \ \ \ .
\end{eqnarray}
As the engineering dimensions of the $\Sigma$ field of 
mesons is zero, one finds that the corresponding quantity 
for the mesonic part of the chiral lagrangian is
\begin{eqnarray}
\partial_\mu s^\mu & = & 
-{f^2\over 4} Tr\left[ \partial_\mu\Sigma \partial^\mu\Sigma^\dagger \right]
\ -\ 
4 \lambda Tr\left[ m_q\Sigma \ +\ h.c \right]
\ +\ ...
\ \ \ \  ,
\end{eqnarray}
where the dots denote terms higher order in the chiral expansion.

In order to complete the relation between 
$ G^{\alpha\beta A} G_{\alpha\beta}^A$
and operators in terms of the $\Sigma$ field, matrix elements 
of the light quark mass operator are required.  
Fortunately, this is straightforward, 
\begin{eqnarray}
\sum_i  m_i\overline{q}_i q_i
& \rightarrow &
-\lambda Tr\left[ m_q\Sigma \ +\ h.c \right]
\ +\ ...
\ \ \ \ ,
\end{eqnarray}
and one finds that 
\begin{eqnarray}
{1\over 2} {\beta(g)\over g} G^{\mu\nu A} G_{\mu\nu}^A 
&\rightarrow &
-{f^2\over 4} Tr\left[ \partial_\mu\Sigma 
\partial^\mu\Sigma^\dagger \right]
\ -\ 
3 \lambda Tr\left[ m_q\Sigma \ +\ h.c \right]
\ +\ ...
\nonumber\\
&\rightarrow & 
-2 \partial_\mu\pi^+\partial^\mu\pi^-\ +\ 3 m_\pi^2 \pi^+\pi^-
\ 
- (\partial_\mu\pi^0)^2\ +\ {3\over 2} m_\pi^2 (\pi^0)^2
\ +\ ...
\ \ \ \  .
\end{eqnarray}
The dots denote higher dimension operators.
The matrix element
$\langle\pi(q^\prime)| 
{1\over 2} {\beta(g)\over g} G^{\mu\nu A} G_{\mu\nu}^A |\pi(q)\rangle $
at tree-level is therefore 
$ m_\pi^2 + p^2$ where $p = q-q^\prime$ is the momentum 
transfer of the $\pi$'s.
This result disagrees with 
\cite{Khara} but agrees with \cite{NovShif}.

We can compute the matrix element to one-loop level 
in the chiral lagrangian.
The higher order counterterms that appear as $dots$ 
in the previous expression
must be retained and 
it is straightforward to show that
\begin{eqnarray}
{\cal A}_G 
& = & \langle\pi(q^\prime)| 
{1\over 2} {\beta(g)\over g} G^{\mu\nu A} G_{\mu\nu}^A |\pi(q)\rangle
\nonumber\\
& \rightarrow  & 
m_\pi^2 + p^2
\ +\ 
\nonumber\\
& & {1 \over 16\pi^2 f^2}
\left[ 
-{31\over 9} m_\pi^2 p^2
\ +\  (3 m_\pi^2 - 2 p^2) p^2 \log\left({m_\pi^2\over\mu^2}\right)
\ +\ \left( {13\over 3}m_\pi^4 + {2\over 3} m_\pi^2 p^2 - 2 p^4\right) 
\Phi_1(p^2/m_\pi^2)
\right.
\nonumber\\
 &   & \left.  \ \ \ \ \ \ -\ 10 m_\pi^2  p^2 \Phi_2 (p^2/m_\pi^2)
 +\  w_1(\mu) m_\pi^2 p^2  + w_2(\mu) p^4 \right]
\ \ \ \ ,
\end{eqnarray}
where the functions $\Phi_{1,2} (Z)$ are defined by 
\begin{eqnarray}
\Phi_1 (Z) & = & \int_0^1\ dx\ 
\log\left( 1-x(1-x) Z - i \epsilon\right)
\nonumber\\
\Phi_2 (Z) & = & \int_0^1\ dx\ x(1-x) 
\log\left( 1-x(1-x) Z - i \epsilon\right)
\ \ \ .
\end{eqnarray}
The constants $w_{1,2}$ are local counterterms arising in 
the chiral lagrangian
whose $\mu$ dependence exactly cancels the $\mu$ dependence 
of the chiral logarithms.
It is natural to choose $\mu = \Lambda_\chi$, 
the scale of chiral symmetry breaking.
We have explicitly retained analytic terms arising from the 
one-loop graph, and have not absorbed them into the counterterms.
Since the contribution of $w_1$ is always suppressed by factors 
of $m_\pi^2/16\pi^2 f^2$
compared to the tree-level amplitude, we will be unable to 
fit $w_1$ to data, 
but we expect that its contribution is  negligible.
It is important to note that the matrix element of 
$G^{\mu\nu A} G_{\mu\nu}^A $ is enhanced by a factor of 
$1/\alpha_s (\mu)$ in its contribution to $\psi\pi$ scattering or
$\pi\pi$  production.

\subsection{ Matrix Elements of $ T_g^{\mu\nu}$}

The matrix element of the traceless part of the gluonic 
component of the 
energy-momentum tensor, $T_g^{\mu\nu}$ is required in 
addition to the 
matrix element of $ G^{\alpha\beta A} G_{\alpha\beta}^A$ 
in order to determine low-momentum scattering amplitudes.
Clearly, we have no symmetry dictating the normalization of 
the matrix elements and we must appeal to data.
The operator $T_g^{\mu\nu}$ is traceless and symmetric 
in its lorentz indices, 
and as such we match onto
an operator in the chiral lagrangian at leading order 
with the same  properties,
\begin{eqnarray}
T_g^{\mu\nu} &\rightarrow &
V_2(\mu) {f^2\over 8} 
Tr\left[  \partial^\mu\Sigma\partial^\nu\Sigma^\dagger
\ +\ \partial^\nu\Sigma\partial^\mu\Sigma^\dagger
\ -\ {1\over 2} g^{\mu\nu} \partial^\alpha
\Sigma\partial_\alpha\Sigma^\dagger
\right]
\ \ \ ,
\end{eqnarray}
where $V_2(\mu)$ is a (scale-dependent) constant 
that must be fit to data.
At tree-level in the chiral expansion the matrix 
element of $T_g^{\mu\nu}$ is
\begin{eqnarray}
{\cal A}_T^{\mu\nu}& = &  
\langle\pi(q^\prime)| T_g^{\mu\nu} |\pi(q)\rangle
\nonumber\\
& = & V_2(\mu)\left[
q^\mu q^{\prime\nu} + q^\nu q^{\prime\mu} 
-{1\over 2} g^{\mu\nu} q\cdot q^\prime
\right]
\  \ \ \ \ ,
\end{eqnarray}
and using the variables $p=q-q^\prime$ and 
$k=(q+q^\prime)/2$ 
\begin{eqnarray}
{\cal A}_T^{\mu\nu}& = & V_2 (\mu)\left[
2\left(k^\mu k^\nu -{1\over 4} g^{\mu\nu} k^2\right)
- {1\over 2}\left(p^\mu p^\nu 
-{1\over 4} g^{\mu\nu} p^2\right)
\right]
\  \ \ \ \ .
\end{eqnarray}

This matrix element can be computed simply at one-loop in the 
chiral expansion.
As $T_g^{\mu\nu}$ is traceless in four-dimensions only, 
we find a contribution
proportional to $g^{\mu\nu}$ that yields a non-zero 
trace at one-loop level.
This contribution is analytic in the external variables 
and masses and is accompanied 
by contributions from local counterterms that are unknown.
We find that 
\begin{eqnarray}
{\cal A}_T^{\mu\nu}& = & V_2 (\mu)
\left[
2\left(k^\mu k^\nu -{1\over 4} g^{\mu\nu} k^2\right)
\right. \nonumber\\
&  & \left.
+ \left( - {1\over 2}\ 
+\  {1\over 8\pi^2 f^2}(2p^2-m_\pi^2)
\left( {1\over 6}\log\left({m_\pi^2\over\mu^2}\right)
\ +\ \Phi_2(p^2/m_\pi^2)\right)\right)
\left(p^\mu p^\nu -{1\over 4} g^{\mu\nu} p^2\right)
\right. \nonumber\\
&  & \left.
+  {1\over 16\pi^2 f^2}
\left( -{1\over 6}m_\pi^4 + {13\over 12} m_\pi^2 p^2 
-{1\over 6} p^4\right) g^{\mu\nu}
\right.
\nonumber\\
& & \left. \ + \ 
{1\over 16\pi^2 f^2}\left( w_3(\mu) m_\pi^2
+ w_4(\mu) p^2 \right)
 \left(p^\mu p^\nu -{1\over 4} g^{\mu\nu} p^2\right)
\right. \nonumber\\
& & \left. \ +\ 
{1\over 16\pi^2 f^2}\left( w_5(\mu) m_\pi^4 
+ w_6(\mu) m_\pi^2 p^2 
+ w_7(\mu) p^4 \right)
 g^{\mu\nu}
\right]
\ \ \ .
\end{eqnarray}
The contributions of $w_3$ and $w_5$ are small and 
will not be able to be fixed by the available data.

For the processes considered here the scale-dependent constant
$V_2 (\mu)$ is not directly observable as it always appears
multiplied by $C_E^{[(1S),(2S)]}+C_B^{[(1S),(2S)]}$.   We will use 
$V_2 (\Lambda_\chi) = {1\over 2}$\cite{NovShif,GRVa}, 
however its actual value will not impact this work.
Deviations from this value will be absorbed into the definition
of $C_E^{[(1S),(2S)]}+C_B^{[(1S),(2S)]}$.
When a comparison is attempted between the extracted values of 
$C_E^{[(1S),(2S)]}$ and $C_B^{[(1S),(2S)]}$ with, say, 
the large-$N_c$ limit of QCD, then the precise value of 
$V_2 (\Lambda_\chi)$ will have consequence.

\section{The Strong Decays
$\pss\rightarrow\ps\pi^+\pi^-$ and  
$\uss\rightarrow\us\pi^+\pi^-$ }

The matrix element for strong decays 
$\pss\rightarrow\ps\pi^+\pi^-$ and  
$\uss\rightarrow\us\pi^+\pi^-$ , e.g.  
\begin{eqnarray}
\langle\ps\ \pi^+ (q_+)\ \pi^-(q_-)| {\cal L}_{\rm int.}|\pss\rangle
\ \ \ ,
\end{eqnarray}
can be obtained from
${\cal A}_G$ and ${\cal A}_T^{\mu\nu}$ by crossing
symmetry.
If $q_+$ and $q_-$ denote the outgoing momentum of the 
$\pi^+$ and $\pi^-$ respectively then we have 
$p=q_++q_-$ and $k={1\over 2} (q_+-q_-)$, 
in the expressions for 
${\cal A}_G$ and ${\cal A}_T^{\mu\nu}$.

The  data we use are from decays between states in quarkonia
to a pion pair.
We have chosen to analyze the data for
$\uss\rightarrow\us\pi^+\pi^-$
\cite{argus}
and  
$\pss\rightarrow\ps\pi^+\pi^-$
\cite{himel}
and not the decays from the $\Upsilon (3S)$ \cite{cleoa}
as we expect the multipole expansion to be more 
reliable for transitions
between low-lying states in quarkonia
(for work on strong decays from the $\Upsilon (3S)$ 
see\cite{moxa}).
The data is presented as a normalized differential decay 
distribution plotted against the variable $x$ defined by 
\begin{eqnarray}
x & = &  { \sqrt{s_{\pi\pi}} - 2 m_\pi\over 
M_{\psi (2S)} - M_{J/\psi} - 2 m_\pi}
\ \ \ ,
\end{eqnarray}
for $\pss\rightarrow\ps\pi^+\pi^-$
as shown in fig.1,
and an analogous definition for the decay 
$\uss\rightarrow\us\pi^+\pi^-$, as shown in fig.2.
$\sqrt{s_{\pi\pi}}$ is the invariant mass of the $\pi\pi$ system.
\begin{figure}
\epsfxsize=10cm
\hfil\epsfbox{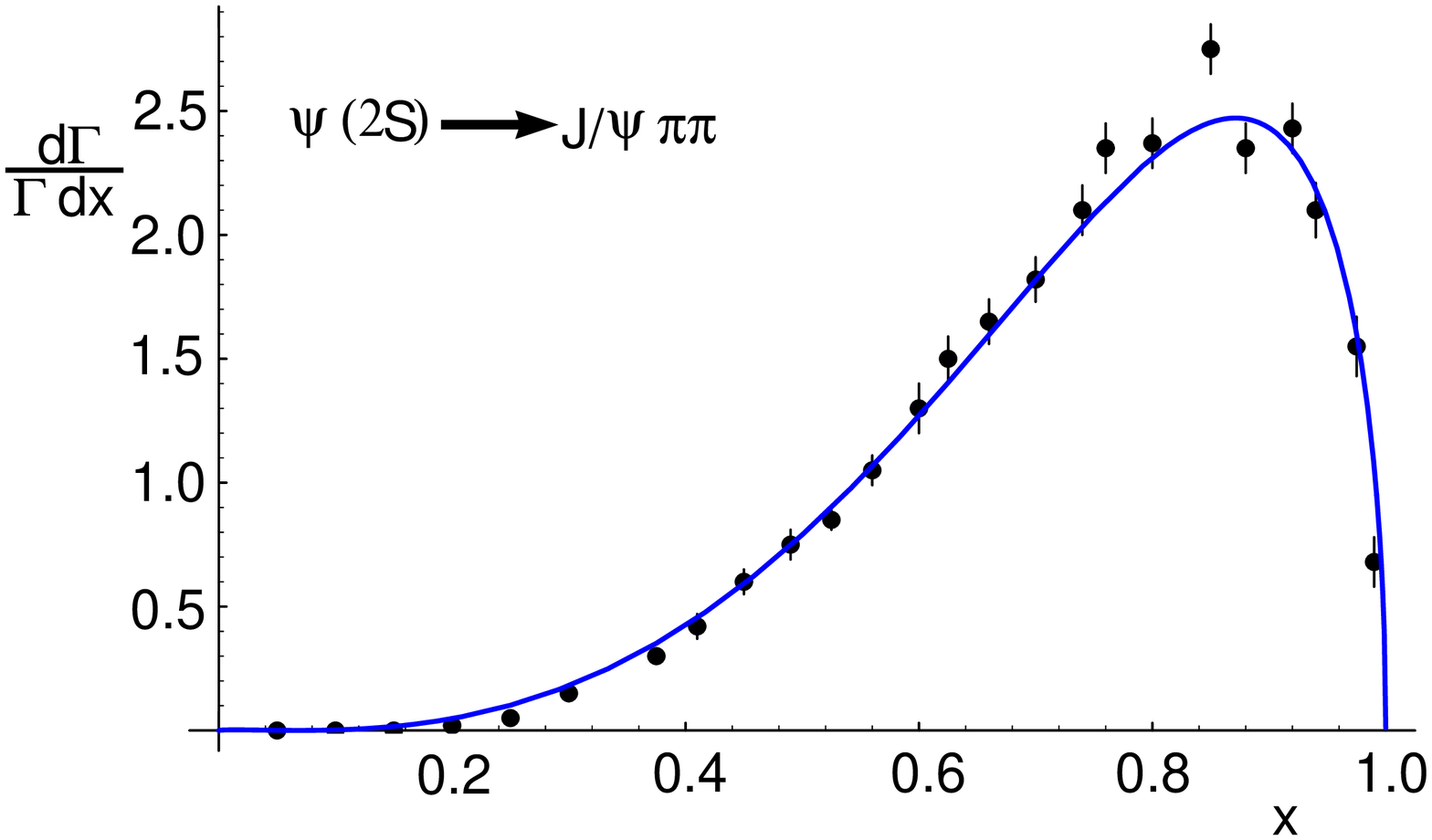}\hfill
\caption{The normalized differential decay spectrum for 
$\pss\rightarrow\ps\pi^+\pi^-$ as function of scaled variable
$x$.
The solid curve is the fit at tree-level.}
\label{psidata}
\end{figure}
\begin{figure}
\epsfxsize=10cm
\hfil\epsfbox{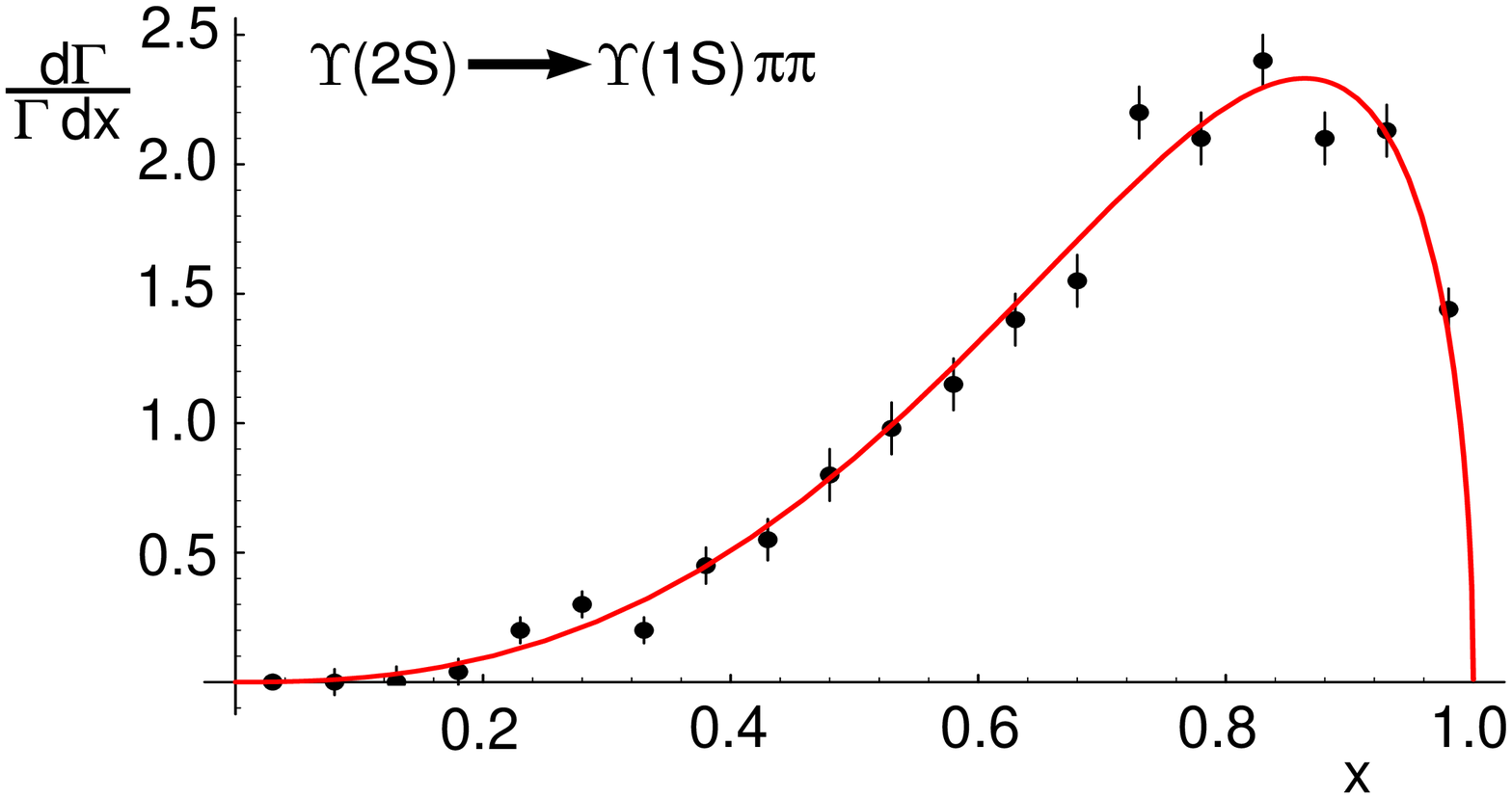}\hfill
\caption{The normalized differential decay spectrum for 
$\uss\rightarrow\us\pi^+\pi^-$ as function of scaled variable
$x$.
The solid curve is the fit at tree-level.}
\label{updata}
\end{figure}
Note that the data we have shown in figs. 1 and 2 is taken 
directly from fig. 7 of 
reference\cite{argus}\  and is not the original data set.

The partial-width for each decay modes is taken from the 
branching fractions and total width measurements 
of the $2S$ states given in 
particle data group evaluation \cite{pdg}
\begin{eqnarray}
\Gamma (\pss\rightarrow\ps\pi^+\pi^-) & = & 90\pm 10 \ {\rm keV}
\nonumber\\
\Gamma (\uss\rightarrow\us\pi^+\pi^-) & = & 8.1\pm 1.3 \ {\rm keV}
\ \ \  .
\end{eqnarray}
In what follows we will use only the central values of the above widths
and neglect the uncertainties.

At leading order in the multipole expansion, and tree-level in the 
chiral lagrangian we can fit the parameters 
$C_E^{[(1S),(2S)]}$ and $C_B^{[(1S),(2S)]}$
(from now on we will drop the superscript, and simply use 
$C_E^{(\psi)} $, $C_B^{(\psi)} $ for the charmonium decays and 
$C_E^{(\Upsilon)} $, $C_B^{(\Upsilon)} $ for the $\Upsilon$ decays)
to the spectra shown in fig.1 and fig. 2.
The tree-level graphs in the chiral lagrangian that contribute to the
decay are shown in fig. 3.
\begin{figure}
\epsfxsize=6cm
\hfil\epsfbox{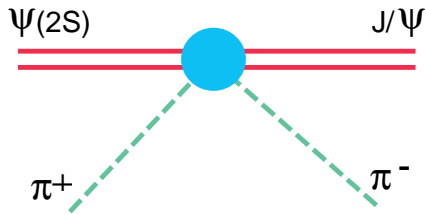}\hfill
\caption{Tree-level diagrams contributing to 
$\pss\rightarrow\ps\pi^+\pi^-$.
The light-shaded circle denotes insertions of
$G^2$ and $T_g^{\mu\nu}$.}
\label{psitreestrong}
\end{figure}
Using a value for the strong coupling of $\alpha_s (r_{c\overline{c}}) = 0.55$
for charmonium and $\alpha_s (r_{b\overline{b}}) = 0.33$ for bottomonium,
we find   ($V_2(\Lambda_\chi)={1\over 2}$)
\begin{eqnarray}
C_E^{(\psi)} - C_B^{(\psi)}  & = & 
7.0\pm 0.2\ {\rm GeV^{-3}} \ \ \ , 
\ \ \ C_E^{(\psi)} + C_B^{(\psi)}   
= 14\pm 1\ {\rm GeV^{-3}}
\nonumber\\
C_E^{(\Upsilon)} - C_B^{(\Upsilon)} & = & 
1.3\pm 0.1\ {\rm GeV^{-3}} \ \ \ , 
\ \ \ C_E^{(\Upsilon)} + C_B^{(\Upsilon)}  
=  3.9\pm 0.7 \ {\rm GeV^{-3}}
\ \ \  .
\end{eqnarray}
In fitting the data we have weighted each data point equally 
(with an error of $\pm 0.1$)  and 
not used the experimental uncertainty indicated by fig.1 and fig.2.
We find a much tighter constraint on the difference 
between $C_B$ and $C_E$ (the coefficient of the $G^2$ operator)
than on the sum (the coefficient of the $T_g^{\mu\nu}$ operator).
These extractions then lead to 
\begin{eqnarray}
C_E^{(\psi)} & = & 10.5\pm 0.6 \ {\rm GeV^{-3}}\ \ \ , 
\ \ \ C_B^{(\psi)} = 3.5\pm 0.5\ {\rm GeV^{-3}}
\nonumber\\
C_E^{(\Upsilon)} & = & 2.6\pm 0.4 \ {\rm GeV^{-3}}\ \ \ , 
\ \ \ C_B^{(\Upsilon)} = 1.3\pm 0.3\ {\rm GeV^{-3}}
\ \ \  .
\end{eqnarray}
We note that in fitting the data there is  a global sign ambiguity
in the coefficients $C_{E,B}$.  We have chosen $C_E > 0$ for our analysis.
The uncertainties 
are from the fit only and do not include the uncertainty associated
with the widths.
\begin{figure}
\epsfxsize=10cm
\hfil\epsfbox{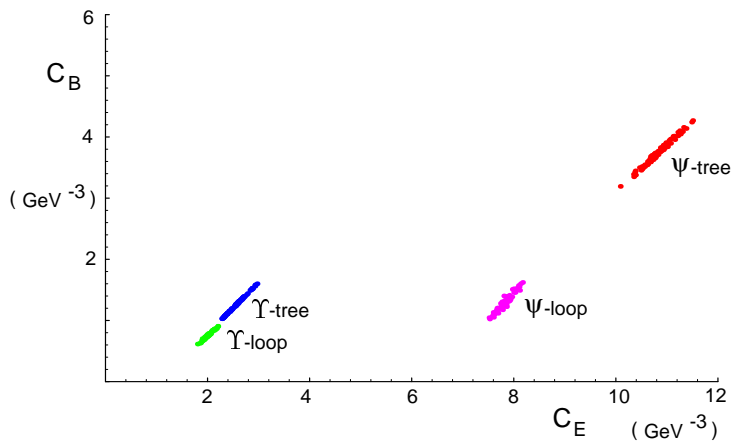}\hfill
\caption{Tree-level  and loop-level 
coefficients $C_E$ and $C_B$ for $\pss-\ps$
and $\uss-\us$ systems.  
The $1-\sigma$ regions arise from a Monte-Carlo 
search in parameter space.
We have used $V_2(\Lambda_\chi)={1\over 2}$, 
$\alpha_s (r_{c\overline{c}}) = 0.55$
and $\alpha_s (r_{b\overline{b}}) = 0.33$.
The loop-level fit is for $w_4=0$.}
\label{treeloopfit}
\end{figure}
We see several interesting trends in these extractions.  
Firstly, the chromo-magnetic coefficient $C_B$ is 
significantly smaller than the 
chromo-electric coefficient $C_E$  for both systems.
This is expected on grounds that the 
magnetic interaction should be suppressed by 
a power of the heavy quark three-velocity in the 
quarkonium.
Secondly,  both coefficients in the $\Upsilon$ system
are substantially smaller than the corresponding quantity in the 
charmonium system.
Again, this is to be expected on the grounds that the 
strong interactions will 
decouple in the limit of vanishing radius of the system.
Our results at this order do not agree with results found 
in \cite{MUa} where coefficients found in that work 
were approximately the same for both systems.

Fitting at one-loop is complicated by the fact that there 
are seven counterterms
$w_1-w_7$.
The counterterms $w_1, w_3, w_5, w_6$ have only a 
small impact upon the 
decays under consideration as they are suppressed by 
powers of the $\pi$ mass 
compared to tree-level terms.
The counterterm $w_7$ occurs in the trace of $T_g^{\mu\nu}$ 
at loop-level and as we have neglected the trace of 
$T_g^{\mu\nu}$ arising at 
loop-level (it is analytic in momenta and masses) 
we have set $w_7=0$.
Naive dimensional analysis suggests that all the counterterms
$w_i$ are of order unity, and hence the effects of setting each $w_i=0$
should be small.   
However, we find this not to be the case.
It is clear from figs. 1 and 2 that the decays are well 
described at tree-level and additional momentum dependence
arising at loop level will be small.
As the amplitudes are dominated by the $G^2$ operator, we suspect
and find that a non-zero value for the $w_2$ counterterm is required.
With so many counterterms we can do little more than give examples.

As an explicit example, we set $w_4=0$ and have $w_2$ as the only
counterterm to be fit to data.  
If the pion energy distributions became available then we 
would be able to
fit both $w_2$ and $w_4$, but at present this is not possible.
The counterterms $w_1$-$w_7$ are the same for all quarkonia 
systems, and  we therefore simultaneously fit 
$C_E^{(\psi)}, C_B^{(\psi)}, C_E^{(\Upsilon)}, 
C_B^{(\Upsilon)}$ and $w_2$
to the data sets for 
$\pss\rightarrow\ps\pi^+\pi^-$ and 
$\uss\rightarrow\us\pi^+\pi^-$.
The loop-level graphs in the chiral lagrangian that contribute to the
decay are shown in fig. 5.
\begin{figure}
\epsfxsize=10cm
\hfil\epsfbox{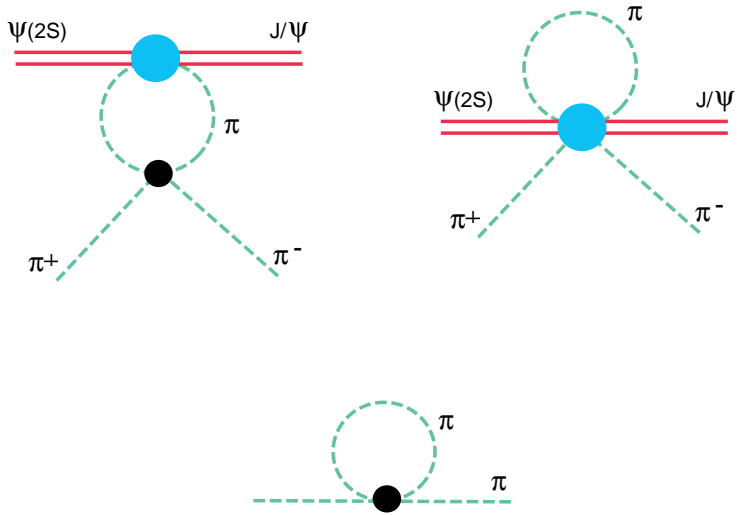}\hfill
\caption{One-loop  diagrams contributing to 
$\pss\rightarrow\ps\pi^+\pi^-$.
The light-shaded circle denotes insertions of
$G^2$ and $T_g^{\mu\nu}$.
The dark circle denotes a strong interaction in the chiral
lagrangian.}
\label{psistrong}
\end{figure}
We find that 
\begin{eqnarray}
C_E^{(\psi)} & = & 7.8\pm 0.4 \ {\rm GeV}^{-3} \ \ \ , 
\ \ \ C_B^{(\psi)} = 1.3\pm 0.3\ {\rm GeV}^{-3} 
\nonumber\\
C_E^{(\Upsilon)} & = & 2.0\pm 0.2\ {\rm GeV}^{-3}  \ \ \ , 
\ \ \ C_B^{(\Upsilon)} = 0.75\pm 0.14\ {\rm GeV}^{-3} 
\nonumber\\
w_2 (\Lambda_\chi) & = & -6.9\pm 0.4
\ \ \  ,
\end{eqnarray}
where the errors on each quantity are correlated.
It is comforting to see that the value of coefficients 
extracted at one-loop level are
only slightly reduced from their tree-level extractions.
One might fear that the relatively large momentum transfer 
involved in these processes
$\sim 590\  {\rm MeV}$ may  make the chiral expansion invalid,
but it would appear that this is in fact not the case.
However, there is a hint of a problem by
the large value of $w_2$. 
The value of $w_2$ is such that it almost cancels the ${\cal O}(p^4)$
contribution from the loop.
It is interesting to note that for this fit we find 
that $C_B$ is much smaller than $C_E$ in the $\psi$-system.
We are able to fit both the $\psi$ and $\Upsilon$ spectra with
single counterterm that is independent of the quarkonium system.
This suggests that higher order terms in the multipole expansion are small.

Fitting both $w_2$ and $w_4$ leads to a slightly better fit but the 
uncertainties associated with the parameters are large.
The central values for the $C_{E,B}$ are very different from the 
tree-level extractions and one might worry that the expansion 
is breaking down, but the large uncertanties preclude any firm 
conclusions.

We note that others have looked at the role of 
final state interactions in these  decays\cite{BDMa}, 
but not using  the framework of chiral perturbation theory.

\section{Comparison with the Large-$N_c$ Limit of QCD}

One is presently unable to match directly from QCD onto the 
lagrange density involving quarkonia and gluons.
Previous estimates of the interaction between 
quarkonia and 
light hadrons have been based largely on 
estimates of the various
$C_E$ and $C_B$ in the large-$N_c$ limit of QCD
\cite{Peska,BhanPesk}.
In the large-$N_c$ limit the repulsion in the 
colour octet channel
vanishes as $1/N_c$
while the attraction in the singlet channel remains.
Consequently, the colour octet intermediate state 
in the dominant 
diagrams in matching onto $C_E$ and $C_B$ can be 
described by plane waves.
Further, the external color singlet states are coulombic.
Matrix elements of a double 
insertion of the chromo-electric dipole operator are simply 
computed in terms of the 
overlap between coulomb bound-state wavefunctions 
and plane waves with 
a plane wave propogator. 
For zero-momentum transfer amplitudes this gives  
a sensible estimate
for the actual matrix element, as there is a mass-gap between the 
intermediate state and the bound-state.
However, for decay processes such as 
$\pss\rightarrow\ps\pi\pi$, the estimate is assured to be 
unreliable as the 
energy release is comparable to the 
gap between the  bound state and the intermediate
state(s).

At leading order there 
is no contribution to $C_B$ while there is a 
calculable contribution to $C_E$.
For dipole-dipole interactions between the same initial and 
final states it is straightforward to show that
\cite{Peska,BhanPesk}
\begin{eqnarray}
C_E^{[(nS),(nS)]} & = & 
{1\over 2} a_0^3 d_n
\nonumber\\
d_n & = & {16\pi\over N_c^2} \ {1\over 3} 
\int {d^3k\over (2\pi)^3}
 {1\over k^2 a_0^2 + \epsilon/\epsilon_0} 
 \left[ { {\bf r^i}\over a_0}\Phi_n\right]^*(k)
 \left[ { {\bf r^i}\over a_0}\Phi_n\right](k)
\ \ \ ,
\end{eqnarray}
(${\bf i}$ is summed over)
where $a_0$ is the bohr radius of the quarkonium 
ground state 
$a_0^\psi\sim (0.7\ {\rm GeV})^{-1}$,
$a_0^\Upsilon\sim (1.5\ {\rm GeV})^{-1}$
\cite{QRa}.
The $\phi_n$ are coulomb wavefunctions for the n-th S-state,
$\epsilon$ is the binding energy of the $n-th$ state and 
$\epsilon_0$ is the binding energy of the ground state.
This calculation leads to 
\begin{eqnarray}
C_E^{[(1S),(1S)]} & = &  a_0^3 {14\pi \over 27}
\ \ \ ,\ \ \ 
C_B^{[(1S),(1S)]} = 0
\nonumber\\
C_E^{[(2S),(2S)]} & = &  a_0^3 {1004\pi \over 27}
\ \ \ ,\ \ \ 
C_B^{[(2S),(2S)]}= 0
\ \ \  ,
\end{eqnarray}
which gives
\begin{eqnarray}
C_E^{[(1S),(1S)]\psi} & \sim & 4.7 \ ({\rm GeV})^{-3}
\ \ \ ,
\ \ \ 
C_E^{[(2S),(2S)]\psi} \sim  341 \ ({\rm GeV})^{-3}
\nonumber\\
C_E^{[(1S),(1S)]\Upsilon} & \sim & 0.48 \ ({\rm GeV})^{-3}
\ \ \ ,
\ \ \ 
C_E^{[(2S),(2S)]\Upsilon} \sim  35 \ ({\rm GeV})^{-3}
\ \  .
\end{eqnarray}

There is a problem in estimating the 
$C_{E,B}^{[(2S),(1S)]} $ from
this construction from the fact that the energy release 
in the transition
is the same magnitude as the off-shellness of the 
intermediate states in the 
free colour-octet propogator.  In this limit a 
short-distance expansion
makes little sense\cite{LSa} and the operator-product 
expansion we have employed is inappropriate.
However, it is more than likely that the colour-octet 
intermediate state 
has a mass splitting to the colour-singlet states, 
although the splitting 
does not become large in any limit of QCD.
The splitting would allow one to perform the OPE as 
we have done and for the 
operators ${\cal O}_{E,B}$ to be the leading terms in a 
systematic expansion.
As such, we cannot directly use the construction of 
the large-$N_c$ limit of QCD to make a reliable 
estimate of the 
$C_{E,B}^{[(2S),(1S)]} $.  
We can compare numerically the values of 
$C_{E,B}^{[(2S),(1S)]} $
we have obtained in previous sections and look 
to see if at 
least they are consistent.
If we naively assume 
\begin{eqnarray}
C_{E,B}^{[(2S),(1S)]} & \sim  &
\sqrt{ C_{E,B}^{[(1S),(1S)]}C_{E,B}^{[(2S),(2S)]} }
\ \ \ ,
\end{eqnarray}
to provide an extremely rough estimate of what 
we might expect for the 
off-diagonal element, then we find that 
$C_{E}^{[(2S),(1S)]\psi}\sim 40  ({\rm GeV})^{-3}$ and 
$C_{E}^{[(2S),(1S)]\Upsilon}\sim 4  ({\rm GeV})^{-3}$
in the large-$N_c$ limit.
These estimates are within factors of 
$\sim 4$ and $\sim 2$ of our extraction.
From this we very cautiously hope that the large-$N_c$ 
estimates of the coefficients
$C_{E,B}^{[(nS),(nS)]}$ are reasonably close to their 
true values
and that the estimates for the binding energy of 
quarkonia in nuclear matter\cite{LMSa} are 
not unreasonable.

\section{Applications}

\subsection{Inelastic $\pi\ps\rightarrow\pi\pss$ and 
$\pi\us\rightarrow\pi\uss$ Scattering}

With the possibility of ``$\ps$-suppression'' being 
a tool
for detecting the formation of a quark-gluon plasma in 
ultra-relativistic heavy ion collisions
it is important to understand inelastic $\ps$ scattering 
in a gas of hadrons.
Naively, one expects $\pi$'s to comprise a non-negligible fraction 
of the number density in a thermal
hadronic gas, and as such, $\pi\ps$ inelastic scattering needs 
to be understood as a background to other mechanisms
that remove the $\ps$'s produced in any given collision.

For $\pi\ps\rightarrow\pi\pss$ and $\pi\us\rightarrow\pi\uss$ 
near threshold ($E_\pi^{\rm lab} \sim 810\  {\rm MeV}$) we can use 
directly the expression for ${\cal A}_G$ and ${\cal A}_T^{\mu\nu}$
discussed previously.
It is likely that there will be significant corrections 
to the amplitudes
for $\pi\ps\rightarrow\pi\pss$ and 
$\pi\us\rightarrow\pi\uss$
determined from the crossed channels
$\uss\rightarrow\us\pi^+\pi^-$ and  
$\pss\rightarrow\ps\pi^+\pi^-$ 
simply due to the large energy transfer involved and 
a break-down of the chiral expansion in this kinematic regime.
We expect that the calculation will fail close to threshold
and
the difference between the cross-section as computed 
at tree-level and 
at one-loop level will give some feeling for the range 
of validity of the 
chiral expansion.
\begin{figure}
\epsfxsize=10cm
\hfil\epsfbox{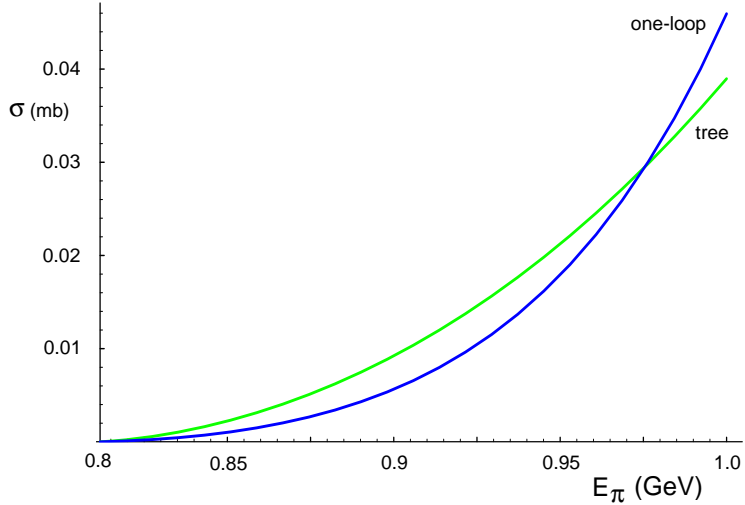}\hfill
\caption{The total cross-section in millibarns for
$\pi\ps\rightarrow\pi\pss$ as  a function of the 
$\pi$ lab energy in GeV.
The curves are evaluated at the value of parameters 
corresponding to the
best fit to $\pss\rightarrow\ps\pi^+\pi^-$
at tree-level and loop-level 
(with $w_2=-6.9$ and the remaining $w_i=0$). }
\label{pipsiscatt}
\end{figure}
The curves in fig. 6 indicate that the loop-level 
and tree-level estimates 
are not so different even near where the expansion must fail.   
This should be taken with extreme caution.
However, we see that the cross section is of the order 
of $10^{-2}$ mb in the 
region where the computation is most reliable.

\subsection{Radiative Decays $\pss\rightarrow\ps\pi^+\pi^-\gamma$ and  
$\uss\rightarrow\us\pi^+\pi^-\gamma$ }

The radiative decays $\pss\rightarrow\ps\pi^+\pi^-\gamma$ and  
$\uss\rightarrow\us\pi^+\pi^-\gamma$ are dominated 
by electromagnetic
interactions of the $\pi$'s that are calculable 
with the chiral lagrangian outlined in the previous sections,
after gauging with respect to the electromagnetic interaction.
Operators involving the electromagnetic field-strength 
tensor $F^{\mu\nu}$ arising 
in the matching between QCD and the effective theory of quarkonium 
will be suppressed by two powers of the quarkonium radius.  
There will be higher dimension operators in the chiral lagrangian 
arising from matching the theory of quarkonia and 
gluons onto the chiral 
lagrangian and quarkonia.
Such operators are suppressed by additional powers of the 
chiral symmetry breaking scale.
The leading contributions to 
$\pss\rightarrow\ps\pi^+\pi^-\gamma$ and  
$\uss\rightarrow\us\pi^+\pi^-\gamma$
occur at tree-level and hence we do not need to discuss the 
role of local counterterms at leading order.
We note that the decays 
$\pss\rightarrow\ps\pi^0\pi^0\gamma$ and  
$\uss\rightarrow\us\pi^0\pi^0\gamma$
are forbidden by charge conjugation.

One might naively expect a significant contribution to the
$\pi\pi\gamma$ decays from pole graphs with  $\chi_{c0} (1P)$,
$\chi_{c1} (1P)$ and  $\chi_{c2} (1P)$ as intermediate states.
However, G-parity forbids the 
$\chi_{cn}(1P)\rightarrow \ps \pi\pi$ transitions and therefore,
such contributions are not present.

It is straightforward to determine the tree-level matrix 
element for the
radiative decays.
There are contributions from both the $G^2$ operator 
and the $T_g^{\mu\nu}$ operator
as shown in fig.7.
\begin{figure}
\epsfxsize=10cm
\hfil\epsfbox{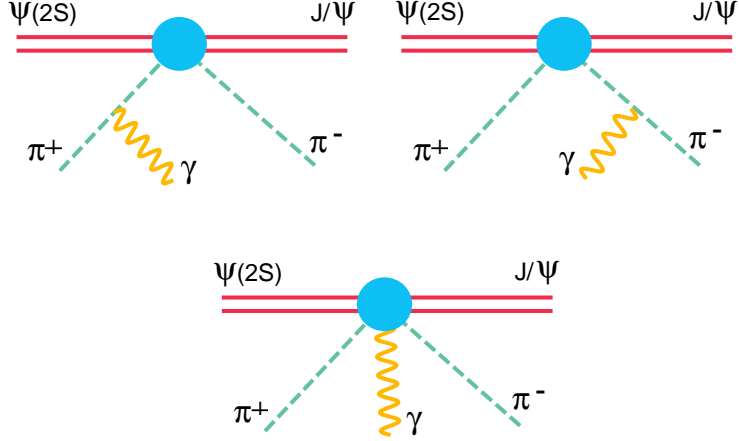}\hfill
\caption{The tree-level diagrams contributing to 
$\pss\rightarrow\ps\pi^+\pi^-\gamma$.
The light shaded circle denotes insertions of the 
$G^2$ and $T_g^{\mu\nu}$ operators.}
\label{psigam}
\end{figure}
The $G^2$ operator yields an amplitude of 
\begin{eqnarray}
i{\cal R}_G & = & 
i e \left( 3 m_\pi^2 + 2 p_+\cdot p_- 
+ 2 k\cdot (p_++p_-)\right) 
\left( {p_+\cdot\varepsilon\over k\cdot p_+} 
- {p_-\cdot\varepsilon\over k\cdot p_-}
\right) 
\ \ \ \ ,
\end{eqnarray}
where $k$ is the photon momentum, $p_{+,-}$ is the 
$\pi^{+,-}$ momentum resp., and 
$\varepsilon$ is the photon polarization tensor.
The contribution from $T_g^{\mu\nu}$ is found to be
\begin{eqnarray}
i{\cal R}_T & = & 
-2 i e V_2(\mu) 
\left[  
v\cdot p_- 
\left( {v\cdot k p_+\cdot\varepsilon\over p_+\cdot k} 
- v\cdot\varepsilon\right)
- 
v\cdot p_+ 
\left( {v\cdot k p_-\cdot\varepsilon\over p_-\cdot k} 
- v\cdot\varepsilon\right)
\right. \nonumber\\
& & \left. + 
\left( v\cdot p_+ v\cdot p_- 
- {1\over 4} p_+\cdot p_- - {1\over 4} k\cdot (p_++p_-) \right) 
\left( {p_+\cdot\varepsilon\over k\cdot p_+} 
- {p_-\cdot\varepsilon\over k\cdot p_-}
\right) 
\right]
\ \ \ \ .
\end{eqnarray}
The full amplitude at this order is given by 
\begin{eqnarray}
i{\cal A}_\gamma & = & 
(C_E+C_B) \left[ i{\cal R}_T \right] 
\ +\ (C_B-C_E) {g\over 2\beta (g) } \left[ i{\cal R}_G \right]
\ \ \ \ .
\end{eqnarray}

The distribution $d\Gamma/dY$ 
resulting from $i{\cal A}_\gamma$
as  a function 
of the normalized photon energy 
$Y=E_\gamma/E_\gamma^{\rm max.}$ 
($E_\gamma^{\rm max.} = 309\ {\rm MeV}$)
is shown for 
different values of the coefficients $C_E$ and 
$C_B$ in fig.8.
In fig.9 we have shown the same differential 
spectrum weighted with
the square of the normalized photon energy 
to highlight the 
higher energy part of the distribution.
Experimentally verifying the differential 
distribution will be a good
test of the chiral and multipole 
expansions.
\begin{figure}
\epsfxsize=10cm
\hfil\epsfbox{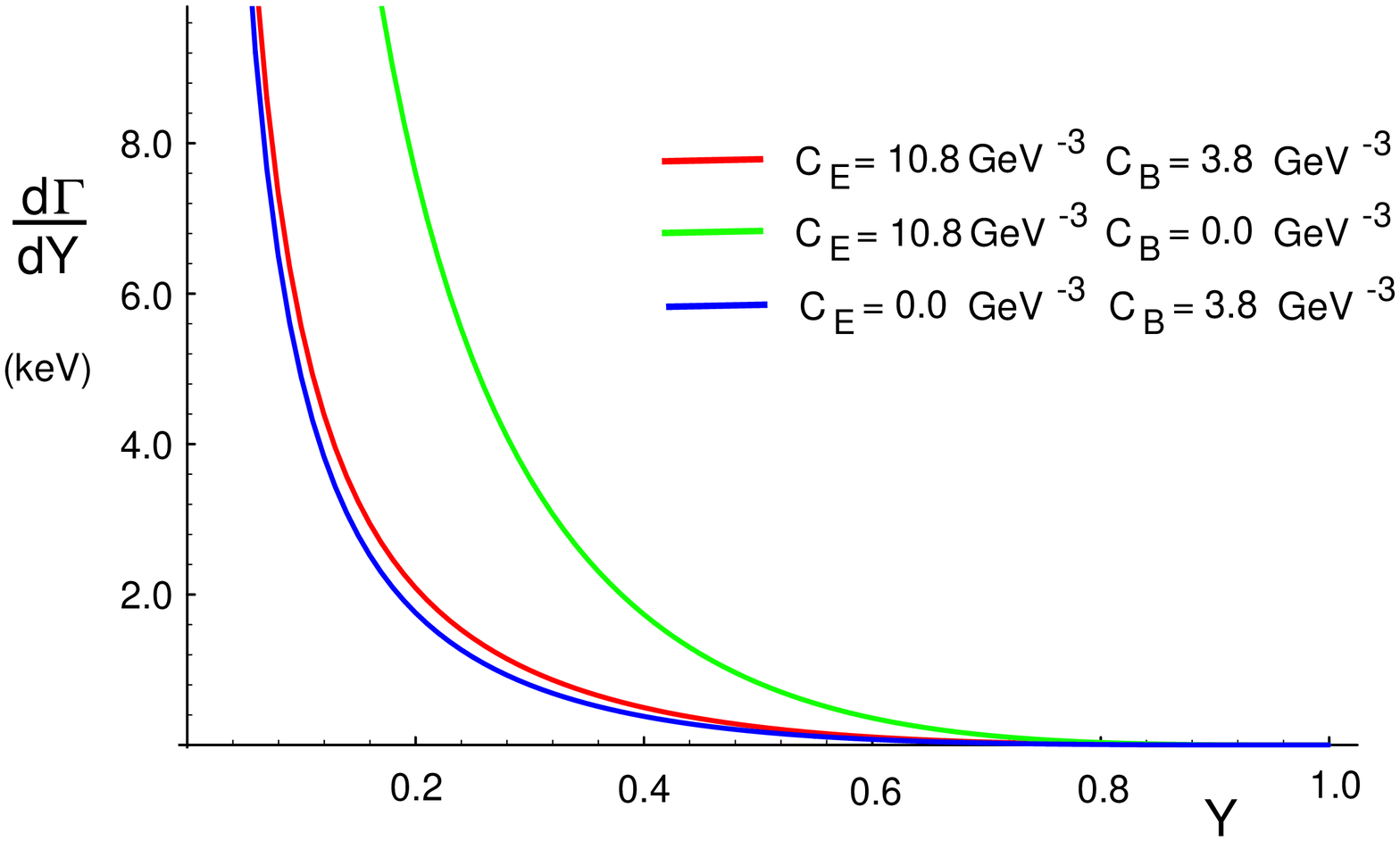}\hfill
\caption{Differential decay spectrum as a 
function of normalized photon energy in
$\pss\rightarrow\ps\pi^+\pi^-\gamma$.
Shown are the distributions obtained for different
values of $C_E$ and $C_B$.
}
\label{psigamdist}
\end{figure}
\begin{figure}
\epsfxsize=10cm
\hfil\epsfbox{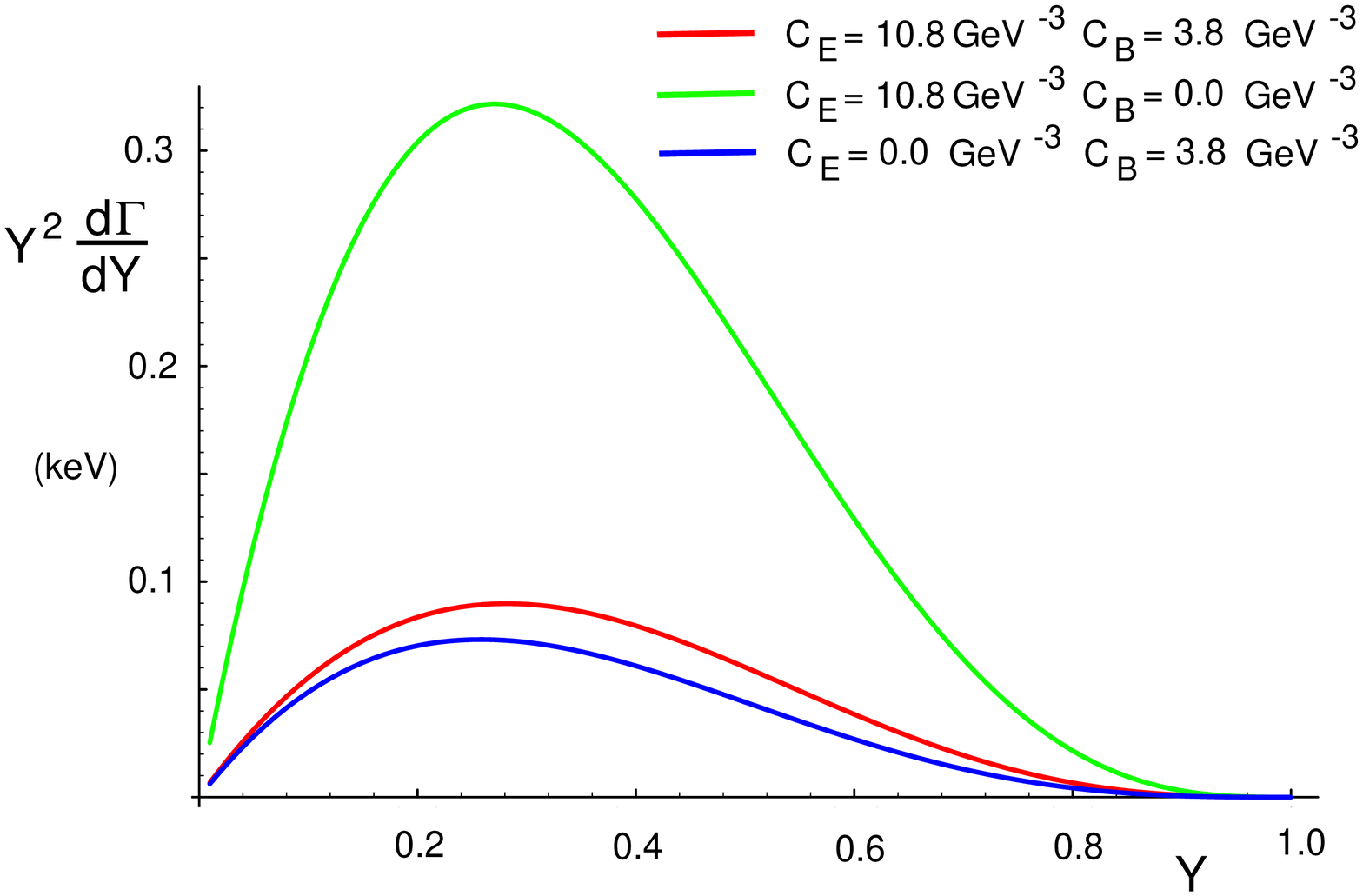}\hfill
\caption{Differential decay spectrum  as a 
function of normalized photon energy in
$\pss\rightarrow\ps\pi^+\pi^-\gamma$ weighted with 
$Y^2$.  
Shown are the distributions for different
values of $C_E$ and $C_B$.}
\label{psigamdistEE}
\end{figure}

The integrated branching fraction is infrared 
divergent as we have 
not included one-loop
vertex and wavefunction graphs. However, it is 
useful to have the 
branching fraction
for events with a $E_\gamma$ greater than some lower 
energy $E_\gamma^{\rm min}$.
\begin{table}
\begin{tabular}{ccc} 
 $Y_{\rm min}$  & {\em Branching Fraction } 
\\   \tableline
\rule{0cm}{0.5cm}
0.2
&$1.0\times 10^{-3}$	
\\  \tableline
\rule{0cm}{0.5cm}
$0.4$
&$2.2\times 10^{-4}$
\\  \tableline
\rule{0cm}{0.5cm}
$0.6$ 
&$3.5\times 10^{-5}$
\\  \tableline
\rule{0cm}{0.5cm}
$0.8$
&$1.8\times 10^{-6}$
\\
\end{tabular}
\vskip 0.5cm
\caption{Branching fractions for 
$\pss\rightarrow\ps\pi^+\pi^-\gamma$
for different cuts on the 
normalized photon energy, $Y_{\rm min}$,
for $\alpha_s(r_{c\overline{c}})=0.55$, 
$C_E=10.8\ {\rm GeV}^{-3} $ 
and $C_B = 3.8\ {\rm GeV}^{-3}$.}
\end{table}
In table 1 we have given the branching fraction for 
different values of $Y_{\rm min}$ for the 
tree-level extracted values of $C_E=10.8\ {\rm GeV}^{-3} $ 
and $C_B = 3.8\ {\rm GeV}^{-3}$.

\subsection{Electromagnetic Polarizability of Quarkonia}

While it is more than likely that the 
electromagnetic polarizability
of quarkonia will never be measured,  there is an 
interesting theoretical
point to be discussed.
The leading contribution to both the electric and magnetic 
polarizabilities,
defined by an effective lagrange density,
\begin{eqnarray}
{\cal L}^{\rm eff.} & = & 
\sum_v\  
\left( P^\dagger_v (f) P_v (i)  
- V^\dagger_{\mu v}(f)  V^\mu_v (i) \right)
\left[ -2 \pi (\alpha+\beta) v_\sigma v_\rho 
F^{\sigma\delta} F^\rho_\delta
\ +\ 
\pi \beta F^{\sigma\delta} F_{\sigma\delta}\right]
\ \ \ ,
\end{eqnarray}
comes from tree-level pole graphs, where two photons 
are coupled to the quarkonium via colour-singlet intermediate states.
The constants $\alpha$ and $\beta$ are the electric and magnetic 
susceptibilities.
Such contributions are finite in the chiral limit and are 
naively estimated to be
\begin{eqnarray}
\alpha\sim\beta & \sim &  
{\alpha_e\over\pi\alpha_s(r_{Q\overline{Q}}) } r_{Q\overline{Q}}^3
\ \ \ ,
\end{eqnarray}
in the same way that $C_{E,B}$ are estimated in the 
large-$N_c$ limit.
However, there is a contribution to $\alpha$ and $\beta$ 
from $C_E$ and $C_B$ 
via $\pi$ loops coupled to two photons, as shown in fig. 10.
\begin{figure}
\epsfxsize=7cm
\hfil\epsfbox{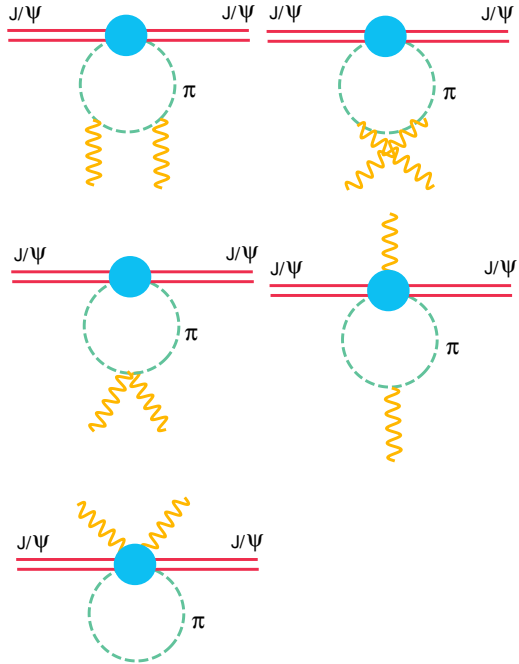}\hfill
\caption{Diagrams contributing to the electromagnetic 
polarizabilities of 
the $\ps$.  The shaded circle denotes insertions of 
$G^2$ and $T_g^{\mu\nu}$.
}
\label{psipolfig}
\end{figure}
This is numerically subleading to the contribution 
from the pole graphs, 
except in the extreme chiral limit.   
In this limit the loop graphs are 
logarithmically infrared divergent, 
\begin{eqnarray}
\alpha^{\rm loop} & = &  \beta^{\rm loop}  = 
-{1\over 12\pi} V_2(\mu) (C_E+C_B) {\alpha_e\over4 \pi}
\log\left({m_\pi^2\over\Lambda_\chi^2}\right) 
\ \ \ ,
\end{eqnarray}
and only $T_g^{\mu\nu}$ contributes.
Using the large-$N_c$ estimates of $C_E=4.7 \ {\rm GeV}^{-3}$ and 
$C_B=0\  {\rm GeV}^{-3}$ along with the actual 
value of the 
$\pi$ mass and $V_2(\mu)={1\over 2}$, 
one finds a loop contribution of
$\alpha^{\rm loop} = \beta^{\rm loop} 
\sim  10^{-6}\ {\rm fm^3}$.

\subsection{ Decays from the $\usss$}

Until this point we have been considering decays from the 
$2S$ states in the $\psi$ and $\Upsilon$ systems.  
However, there is data available for the transitions
$\usss\rightarrow\us\pi\pi$ as shown in fig. 11, and 
$\usss\rightarrow\uss\pi\pi$ \cite{cleoa}.
\begin{figure}
\epsfxsize=10cm
\hfil\epsfbox{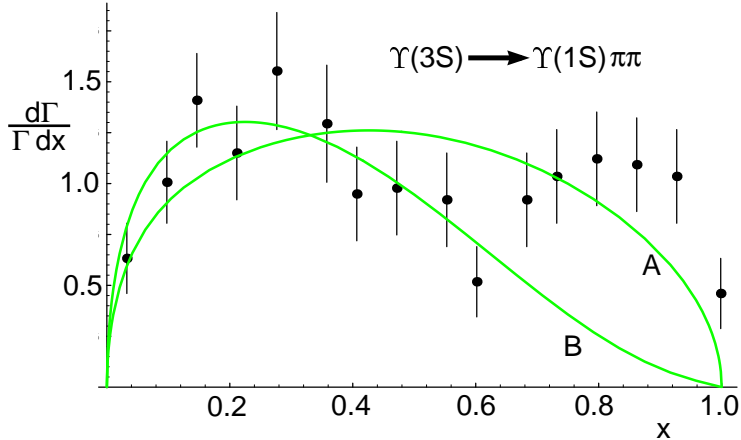}\hfill
\caption{Spectrum for $\usss\rightarrow\us\pi^+\pi^-$ verses $x$.
The solid curve {\bf B} shows the best tree-level fit for the data
with $m_{\pi\pi} < 0.65 {\rm GeV}$, while curve {\bf A} is the 
best fit over all $x$.}
\label{updata3Sx}
\end{figure}
(Note that we have taken the data from fig. 11 of \cite{cleoa} 
and it is not the original data set.)
The spectrum for $\usss\rightarrow\uss\pi\pi$ is 
similar to that of the 
decays from the $2S$ states and we will not discuss it further.
There has been significant discussion of 
$\usss\rightarrow\us\pi\pi$  as the 
differential spectrum does not have the same shape as that of
transitions from the $2S$ states
\cite{BDMa,Volres,LTa}.
It has been suggested that a resonance in the $\us\pi$ channel
\cite{BDMa,Volres} is responsible for the observed distribution.
Another suggestion was that the multipole expansion is subdominant
to hadronic effects involving B-mesons\cite{LTa}.
It is important to note that the invariant mass of the $\pi\pi$ 
system can extend up to $0.9\ {\rm GeV}$.
The soft pion analysis is expected to fail for the 
higher mass pairs.
As such, one might expect to see modifications to 
the differential
spectrum before the end-point.
Further, it was assumed that $C_B$ vanished in the 
previous analyses, 
allowing only for a spectrum peaked near the high 
end of the spectrum.
It is clear from fig. 11 that the best tree-level 
fit to the entire range of $x$ is poor, 
with the apparent minimum at $x\sim 0.6$ not reproduced.
If we restrict ourselves to the low end of the spectrum where the 
chiral lagrangian is expected to be applicable, 
then that portion of the spectrum
can be described with the multipole expansion.
$T_g^{\mu\nu}$ dominates the decay amplitude with 
$C_E\sim C_B\sim 0.3$
yielding the best fit for 
$m_{\pi\pi} < 0.65 {\rm GeV}$.
Clearly, only a portion of the spectrum can be described at 
leading order in the 
multipole and chiral expansions.
Whether the deviations from the leading result arises 
from higher 
order terms in the chiral or multipole expansions or
from a given resonance
cannot be determined.
However, in order to reproduce the low-mass region of 
the spectrum, 
a large value for the chromo-magnetic coupling $C_B$ is required, 
in contrast to transitions
from the $2S$ states.

\section{Discussion}

We have explored the strong interactions and radiative 
decays of quarkonia in order to better understand the 
multipole and chiral expansions that might be appropriate
for such  processes.
The unproven convergence of the multipole expansion is assumed 
throughout this work and we suggest that numerical factors could
lead to convergence despite there being no rigorous limit
of QCD in which this is true.
Current discussions of how quarkonia interact
with the light hadrons away from the chiral limit 
have been addressed and 
the strong decays $\pss\rightarrow\ps\pi^+\pi^-$,
$\uss\rightarrow\us\pi^+\pi^-$ were focused on 
at tree-level 
and loop-level in the chiral expansion.
The leading couplings in the multipole expansion are 
found to be significantly smaller in the $\Upsilon$ 
system than in the $\psi$ system, consistent with 
naive scaling arguments.
Further, counterterms appearing at loop level in the chiral expansion
are found to be approximately
the same for the $\psi$ and $\Upsilon$ systems, 
suggesting that the multipole expansion is converging.
In addition,  we discussed the large-$N_C$ limit in 
which estimates of the 
coefficients $C_{E,B}$ for various quarkonium states have been made.
While there is nothing concrete that can be concluded, 
it would appear that the large-$N_C$ limit  is not in serious
conflict with data.
Crossing symmetry has been used to estimate the  size of the 
inelastic $\pi\ps\rightarrow\pi\pss$ scattering cross section
that might be relevant for understanding $\psi$ 
suppression in ultra-relativistic heavy ion collisions.  
In the region where the theory is expected to be most 
reliable we find a small cross section, $\sim 10^{-2}\ {\rm mb}$.
Power counting in the quarkonium radius indicates that the 
radiative decays $\pss\rightarrow\ps\pi^+\pi^-\gamma$, 
$\uss\rightarrow\us\pi^+\pi^-\gamma$ are dominated by 
tree graphs computable in chiral perturbation theory.   
The photon energy spectrum is presented and 
 branching fractions $\sim 10^{-4}$ are found.
These branching fraction depends sensitively upon the 
values of $C_{E,B}$
used and thus provides a  consistency test of the calculational
framework.
Finally, we made brief remarks on transitions from the $\usss$
 and on the electromagnetic polarizability of the $\psi$.

\vfill\eject

\centerline{\bf Acknowledgements}

We would like to thank Jerry Miller for discussions 
that lead to this work.
We also thank M. Luke for his critical reading of the manuscript.
This work is supported by the Department of Energy.

\end{document}